\documentclass[12pt,preprint]{aastex}
\usepackage{amsmath,amssymb}

 \newcommand{\citar}[1]{\citeauthor{#1} (\citeyear{#1})}
\newcommand{\citarwiths}[1]{\citeauthor{#1}'s (\citeyear{#1})} 
  \newcommand{\citarNP}[1]{\citeauthor{#1} \citeyear{#1}}
 
\newcommand{\citartwoNP}[2]{\citeauthor{#1} \citeyear{#1}, \citeyear{#2}} 

\begin{document}

\title{Evidence for collisional depolarization in the MgH lines
of the second solar spectrum}

\author{A. Asensio Ramos\altaffilmark{1} and J. Trujillo Bueno\altaffilmark{1,2}} \altaffiltext{1}{Instituto de Astrof\'{\i}sica
de Canarias, 38205, La Laguna, Tenerife, Spain} \altaffiltext{2}{Consejo Superior de Investigaciones Cient\'{\i}ficas,
Spain} \email{aasensio@iac.es, jtb@iac.es}

\begin{abstract}
Analysis of the Hanle effect in solar molecular lines allows us to obtain empirical information on hidden,
mixed-polarity magnetic fields at subresolution scales in the 
(granular) upflowing regions of the `quiet' solar photosphere. Here we report that
collisions seem to be very efficient in depolarizing the rotational levels of MgH lines. This has the interesting consequence that in
the upflowing regions of the quiet solar photosphere the strength of the hidden magnetic field cannot be sensibly larger than 10 G,
assuming the simplest case of a single valued microturbulent field that fills the entire upflowing photospheric volume.
Alternatively, an equally good theoretical fit to the observed scattering polarization amplitudes can be achieved by assuming that
the rate of depolarizing collisions is an order of magnitude smaller than in the previous collisionally dominated case, but then the
required strength of the hidden field in the upflowing regions turns out to be unrealistically high. These constraints
reinforce our previously obtained conclusion that there is a vast amount of hidden magnetic energy and unsigned magnetic flux
localized in the (intergranular) downflowing regions of the quiet solar photosphere.
\end{abstract}


\section{Introduction}

One of the interesting observational discoveries of the last decade is that several diatomic molecules in the ``quiet'' regions of
the solar photosphere, such as MgH, ${\rm C}_2$ and CN, show conspicuous scattering polarization signals when observing close to the edge of the solar disk (\citarNP{stenflo_keller97}; \citartwoNP{gandorfer_atlas1_00}{gandorfer03}; \citarNP{stenflo03}). Of particular interest is the fact, reported by \citar{gandorfer_atlas1_00} and \citar{stenflo03}, that the molecular scattering polarization amplitudes at a given distance from the solar limb appear to be both spatially invariant and independent of the solar magnetic activity cycle, in sharp contrast with the behavior of many strong atomic lines.

For us, this was indeed a truly puzzling behavior because, as pointed out by \citar{egidio03} and \citar{trujillo_spw3_03} during the third international workshop on solar polarization, the sensitivity of molecular lines to the Hanle effect should actually be
similar to that of atomic lines. It is true that the Land\'e factors ($g_L$) of the molecular line levels are typically much smaller
than those of atomic lines (\citarNP{berdyugina02b}), but this does not imply at all that molecular lines are ``immune'' to the
Hanle effect because one has to take into account that the 
radiative lifetimes ($t_{\rm life}$) of molecular levels are generally larger than those of atomic lines. Consequently, the critical Hanle field strength, $B_H=1.137{\times}10^{-7}\,/\,(t_{\rm life}g_L)$, is
similar for both atomic and molecular lines\footnote{$B_H$ is the field strength that is sufficient to
produce a significant change in the line's 
scattering polarization amplitude when the excitation
of the atomic or molecular system is dominated by radiative transitions. If collisions are also efficient then the critical Hanle field
increases due to collisional quenching.} 
(e.g., $B_H{\approx}23$ G for the Sr {\sc i} 4607 \AA\ line and $B_H{\approx}22$ G for the 5175.38 \AA\ line of MgH). 

The resolution of that puzzling behavior was found when it was pointed out (\citarNP{trujillo_spw3_03}) that the observed scattering
polarization in very weak spectral lines, such as those of molecules, is coming {\em mainly} from the upflowing
regions of the ``quiet'' solar photosphere 
(see Fig. 2 of Trujillo Bueno et al. 2004). 
According to atomic physics, the scattering
polarization in molecular lines is indeed sensitive to the Hanle effect. What happens in the Sun is that the probability density
function (PDF) that describes the distribution of magnetic fields in the (granular) upflowing regions is dominated by very weak fields, so that the magnetic fields that could in principle produce a {\em substantial} Hanle depolarization have a very small filling factor (\citarNP{trujillo_spw3_03}). For instance, if we assume that the shape of that PDF is an exponential we then find $\langle B \rangle \approx 15$ G when applying the Hanle effect line ratio technique for C$_2$ lines reported by \citar{trujillo_spw3_03}, which corresponds to $\langle B \rangle \approx 7$ G for the simpler case of a single valued field (\citarNP{trujillo_nature04})\footnote{This conclusion 
has been confirmed by \citar{berdyugina_fluri04} 
by using three unblended C$_2$ lines that are thought to 
be more sensitive to the weaker fields. 
The value $\langle B \rangle \approx 15$ G
that they reported {\em for the single valued microturbulent field case} is an overestimation, given that their analysis is in error
by a factor of 2 because C$_2$ does not have $\Lambda$-doubling. This fact has been independently noticed by those authors.}.

The main purpose of this letter is to report on the very interesting finding summarized in the abstract. As shown below, our
theoretical investigation is based on three-dimensional (3D) radiative transfer modeling of the scattering polarization in each of the 37
unblended MgH lines that show significant linear polarization amplitudes in \citarwiths{gandorfer_atlas1_00} atlas of the
linearly-polarized solar limb spectrum.

\section{Formulation of the problem}

We have solved the 3D scattering polarization problem in
each unblended MgH line in a way similar to that pursued by \citar{trujillo_nature04} for the Sr {\sc i} 4607 \AA\ line case --that is, by using a realistic 3D model of the solar photosphere resulting from Asplund's et al. (2000) hydrodynamical simulations of solar surface convection.

The MgH lines that produce scattering polarization signals in the Sun are located around 5100 \AA\ and they result from $\Delta J=J_u-J_l=0$ (Q-branch) transitions between the rotational $J_u$-levels of the $v_u=0$ vibrational level of the excited electronic state, ${\rm
A}^2{\Pi}$, and the $J_l$-levels of the $v_l=0$ level of the ground electronic state, ${\rm X}^2{\Sigma}^{+}$. The separation between adjacent $J$-levels within each vibrational level is of the order of
$10^{10}$ s$^{-1}$, which is a few orders of magnitude larger than the level's natural width. It follows that in the absence of magnetic fields quantum interferences between such $J$ levels can be neglected,
which is expected to be a suitable approximation at least up to the magnetic field strength that establishes the transition to the Paschen-Back regime for the very first $J_l$ levels (i.e., at least up to $B{\approx}280$ G according to \citarNP{berdyugina02a}). The Land\'e factors we have used are the exact ones, which for the upper levels correspond to the intermediate coupling scheme between Hund's cases (a) and (b), while for the lower levels to Hund's case (b). Interestingly, because for MgH lines with $J{\ge}4.5$
the Land\'e factor $g_L\,{\propto}\,1/J$, we have that the Hanle field $B_H\,{\propto}\,J$. Thus, for example, $B_H{\approx}13$ G 
for the $Q_1(J=6.5)$ line, while it is $B_H{\approx}78$ G for the $Q_1(J=37.5)$ line. 
It is of interest to point out that, in principle, this differential sensitivity to the Hanle effect may be modified due to the possibility of a dependence with $J$ of the collisional rates. We will address this differential collisional quenching issue in more detail in a forthcoming publication.

Although we have solved the 
radiative transfer equations numerically in the
chosen 3D photospheric model, it is useful
to note that for the limiting case of a tangential 
observation in a plane-parallel atmosphere the emergent
fractional linear polarization is approximately given by

\begin{eqnarray}
\frac{Q}{I} \,{\approx}\, \frac{3}{2\sqrt{2}}\,\,\frac{\eta_I^{l}}{\eta_I^{l}+\eta_I^{c}}\,[\frac{S_l}{S_\nu}\,w^{(2)}_{J_uJ_l}\,\sigma^2_0({J_u})\,-\,w^{(2)}_{J_lJ_u}\,\sigma^2_0({J_l})],
\end{eqnarray}
where $\sigma^2_0({J})={\rho^2_0(J)}/{\rho^0_0(J)}$ is the fractional alignment of the $J$-level (being $\rho^0_0(J)$ and $\rho^2_0(J)$ the multipolar components of the density matrix corresponding to the $J$-level, so that $\rho^0_0(J)$ is proportional to the overall population of the $J$-level while $\rho^2_0(J)$ is non-zero if the population of substates with different values of $|M|$ are different). The rest
of the symbols in Eq. (1) have their usual meaning (see \citarNP{trujillo_spw3_03}).

We have calculated the number density of MgH molecules at each grid point of the 3D photospheric model by applying the instantaneous
chemical equilibrium approximation, which is a reliable one in the solar photosphere (\citarNP{asensio_trujillo_spw3_03}).
Moreover, a two-level molecular model for each particular line transition is a good approximation for calculating $\sigma^2_0({J})$, for both $J=J_u$ and $J=J_l$
(\citarNP{egidio03}; \citarNP{asensio_trujillo_spw3_03}). It is clarifying to note that the fractional polarization of each $J$-level is approximately given by

\begin{eqnarray}
{\sigma}^2_0(J)\,{\approx}\,\, \mathcal H_{J J^{'}}^{(2)}(J)
\frac{J^2_{0}}{J^0_0},
\end{eqnarray}
where ${J^2_{0}}/{J^0_0}$ is the degree of anisotropy of the radiation field at the line's frequency (e.g., \citarNP{trujillo01}), while $\mathcal H_{JJ^{'}}^{(2)}(J)$ (being $J^{'}$ the
total angular momentum of the lower level ``$l$'' if $J$ is that of the upper level ``$u$'', and viceversa) accounts for the joined
depolarizing action of collisions and of the assumed microturbulent field.

If the ground level is assumed to be unpolarized (i.e., $\sigma^2_0({J_l})=0$), then the relevant 
statistical equilibrium equations for the microturbulent field case are those derived by \citar{trujillo_manso99}, which give

\begin{eqnarray}
\mathcal H_{J_u J_l}^{(2)}(J_u)={\cal H}^{(2)}w^{(2)}_{J_uJ_l} {{1-\epsilon}\over{1+{\delta}_u(1-\epsilon)}},
\end{eqnarray}
where ${\cal{H}}^{(2)}$ is the Hanle depolarization factor,
$\delta_u=D_u^{(2)}{\,}t_{\rm life}$ 
quantifies the upper-level rate of elastic (depolarizing) collisions in units of the inverse of the upper-level's lifetime,
and $\epsilon$ is the probability that a de-excitation event is caused by inelastic collisions.

For the general case in which we have population imbalances 
in both levels the problem is much more complicated because it requires solving jointly the rate equation for each $\rho^K_0(J)$ multipolar component of the lower and upper levels assuming statistical steady state and the Stokes-vector transfer equations (see \citarNP{trujillo03}). Fortunately, it can be shown that under the weak anisotropy limit approximation discussed by \citar{landi_landolfi04} a relationship similar to that of Eq. (2) holds for each $J$-level (i.e., for both $J=J_u$ and $J=J_l$), although with
considerably more involved expressions for $\mathcal H_{JJ^{'}}^{(2)}(J)$.

\section{Results}

Here we investigate which combinations of field strengths
and collisional rates produce polarization amplitudes in agreement with \citarwiths{gandorfer_atlas1_00} observations. For
simplicity in the presentation, the selected calculations below correspond to the case in which the same elastic collisional rate is
assumed for both the upper and lower levels (i.e., $D_l^{(2)}=D_u^{(2)}=D^{(2)}$). As will be shown below in Fig. 4, we have done this exercise for each of the 37 unblended MgH lines that show measurable $Q/I$ signals, whose $J$-values span from 6.5 to 37.5. However, we show first details of our calculations for a representative MgH line: the Q$_1(J=10.5)$ line at 5175.3867 \AA, whose Einstein coefficient for spontaneous emission is $A_{ul}=0.99 \times 10^7$ s$^{-1}$ (\citarNP{weck03}).

\subsection{The case without collisions}

\begin{figure}
\plotone{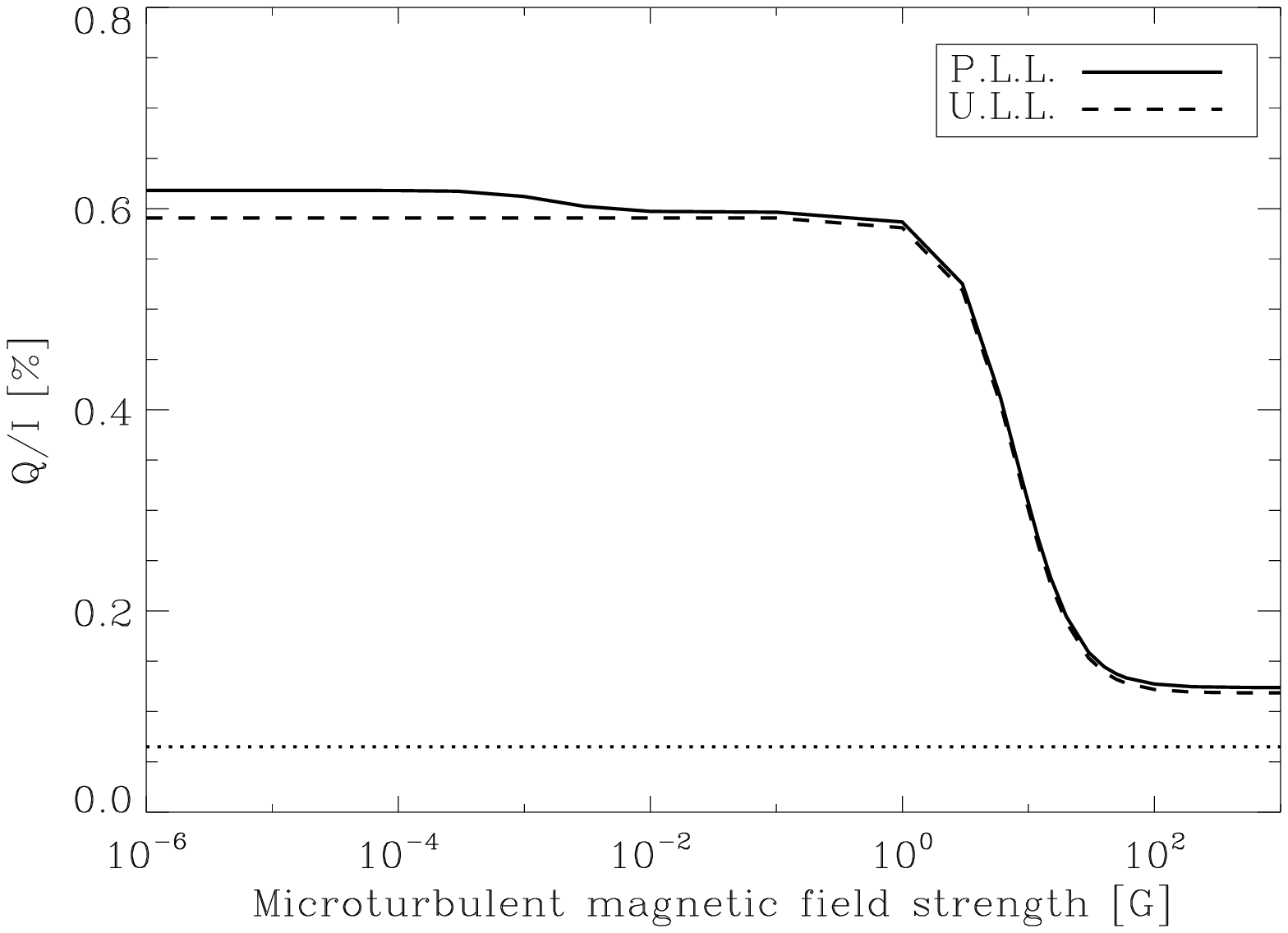}
\caption{The case without collisions. This figure shows the
magnetic sensitivity of the calculated horizontally averaged $Q/I$ signal in the core of the
5175.38 \AA\ MgH line (obtained in the 3D photospheric model for a line of sight at $\mu=\cos {\theta}=0.1$, with ${\theta}$ the
heliocentric angle). The dashed
line refers to the unpolarized lower-level case for which
only selective emission processes contribute to the
emergent linear polarization, while the
solid line corresponds to the most general situation in which the lower level can also be polarized, and both selective 
emission and selective absorption processes 
contribute to $Q/I$. The fact that both
curves are very similar even for field strengths weaker than 10 mG
confirms that it is practically impossible to obtain any observational hint on the presence of ground-level polarization in such solar MgH lines by means of on-disk observations (see \citarNP{trujillo_spw3_03}).
Note that the only action of a volume-filling microturbulent magnetic field is not sufficient to produce the depolarization needed for explaining the observed $Q/I$ amplitude indicated by the
dotted line. \label{fig_qi_field}}
\end{figure}

Firstly, we consider the \emph{collisionless} case in which we have only varied the strength of the microturbulent magnetic field,
which is assumed to have the same value at all the grid-points
of the 3D atmospheric model. Figure \ref{fig_qi_field} shows that magnetic depolarization alone is not sufficient to explain the
observed scattering polarization amplitude.

\subsection{The case without magnetic fields}

\begin{figure}
\plotone{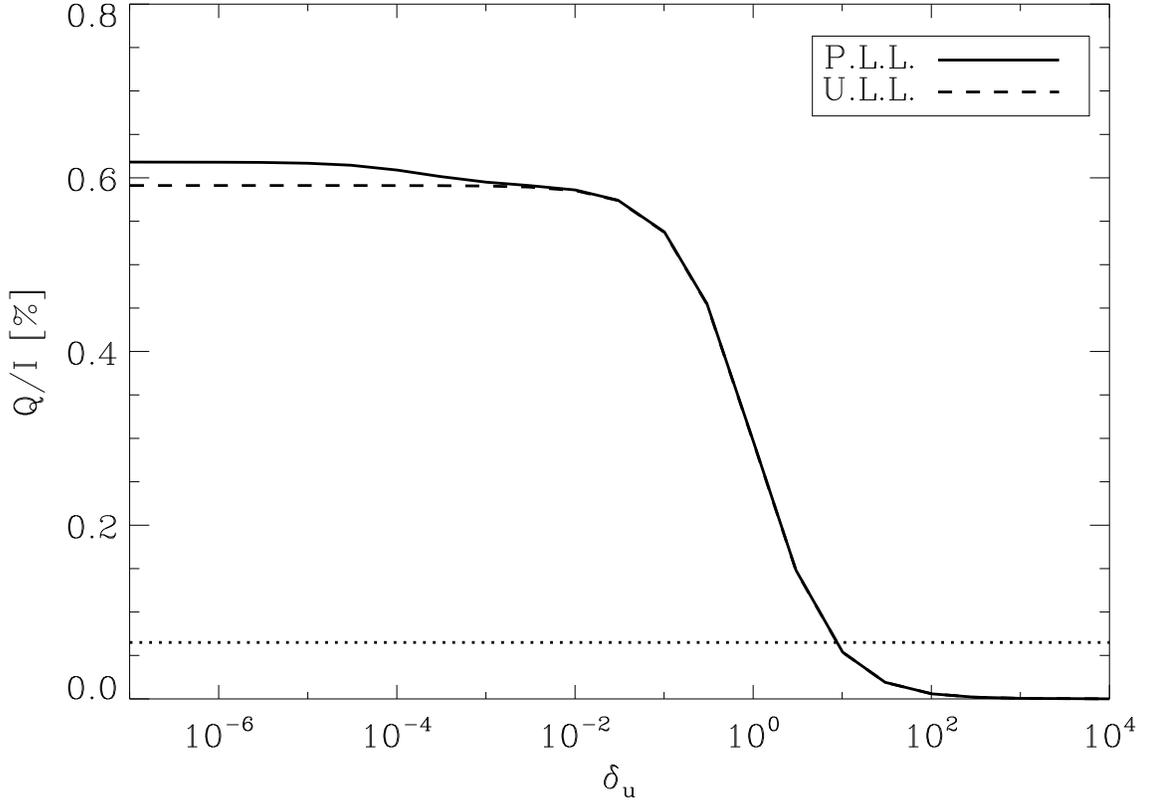}
\caption{The case without magnetic fields. This figure is similar to that of Fig. \ref{fig_qi_field}, but showing instead the variation of $Q/I$ at  the core of the 5175.38 \AA\ MgH line when the elastic
collisional rate is increased. The solid line corresponds to the most general case in which the lower level can be polarized, while
the dashed line refers to the unpolarized lower-level case. 
Note that the elastic collisional rate necessary for obtaining the observed $Q/I$ amplitude 
indicated by the dotted line is very significant -that is, $\delta_u = D^{(2)} {\,} t_{\rm life} {\approx} 9 $, and where $D^{(2)}$ 
can be estimated by using 
$t_{\rm life} {\approx}1/A_{ul}{\approx}10^{-7}{\rm s}$.
\label{fig_qi_delta}}
\end{figure}

Secondly, we consider the case in which only the rate of elastic collisions is increased, while both inelastic collisions and
magnetic fields are disregarded. Figure \ref{fig_qi_delta} shows that the collisional depolarization rate needed for obtaining a $Q/I$
value compatible with \citarwiths{gandorfer_atlas1_00} observations is $\delta_u=D_u^{(2)}{\,}t_{\rm life} {\approx} 9$. This is an order-of-magnitude greater than the $\delta_u$-value
reported by Mohan Rao \& Rangarajan (1999) for the $\lambda$5165.933 line, which they obtained
via a simplistic 1D modeling approach.

\subsection{The case with collisions and magnetic fields}

\begin{figure}
\plotone{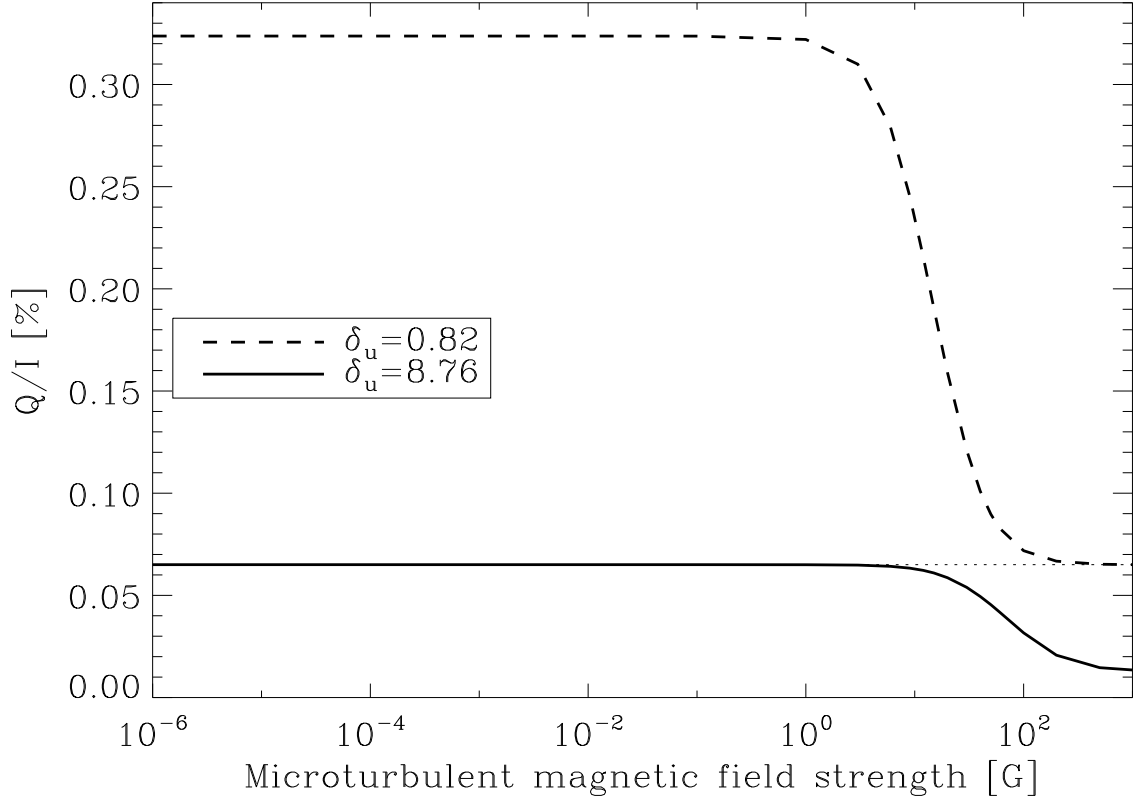}
\caption{This figure is similar to that of Fig.
\ref{fig_qi_field}, but for two ``extreme'' 
non-zero values of the elastic collisional rate.
The solid line shows the magnetic field sensitivity 
of the calculated $Q/I$ signal when the
elastic collisional rate is the one that fits the observed $Q/I$ amplitude in the absence of magnetic fields (i.e.,
$\delta_u=8.76$). The dashed line shows the case in which the elastic collisional rate is $\delta_u=0.82$ (i.e., that which fits 
the observed $Q/I$
amplitude in the presence of a sufficiently strong microturbulent field that produces Hanle effect saturation in both levels). The aim of this figure is to point out that 
in the collisionally dominated case (solid line) the strength of the
microturbulent field cannot be sensibly larger than 10 G, while 
in the weakly collisional case (dashed line) 
it would have to be stronger than 100 G in order to fit the
polarization amplitude observed in the 5175.38 \AA\ MgH line
\label{fig_qi_field_collis}}
\end{figure}

The solid line of Fig. \ref{fig_qi_field_collis} shows the magnetic field sensitivity of the spatially averaged $Q/I$
calculated assuming $\delta_u=8.76$ --that is, using the elastic collisional rate value that in the absence of magnetic fields fits
the $Q/I$ amplitude observed in the 5175.38 \AA\ MgH line. The important point to remember is that in this collisionally dominated
case a good agreement with the observed
scattering polarization amplitude is found only if the assumed microturbulent field is weaker than about 10 G.

The dashed line of Fig. \ref{fig_qi_field_collis} shows what happens when we do the same numerical experiment but using instead $\delta_u=0.82$ --that is, the elastic collisional rate that in the presence of a sufficiently strong
microturbulent field (so as to produce complete Hanle saturation in both levels of the 5175.38 \AA\ MgH line) fits the observed
$Q/I$ amplitude. The main point we want to highlight now is that in this weakly collisional case the minimum strength of the 
microturbulent field needed to fit the observed polarization amplitude
must be {\em at least} $\sim$100 G. In our opinion, this represents an unrealistically high value for the strength of the microturbulent
field in the (granular) upflowing regions.

\begin{figure}
\plotone{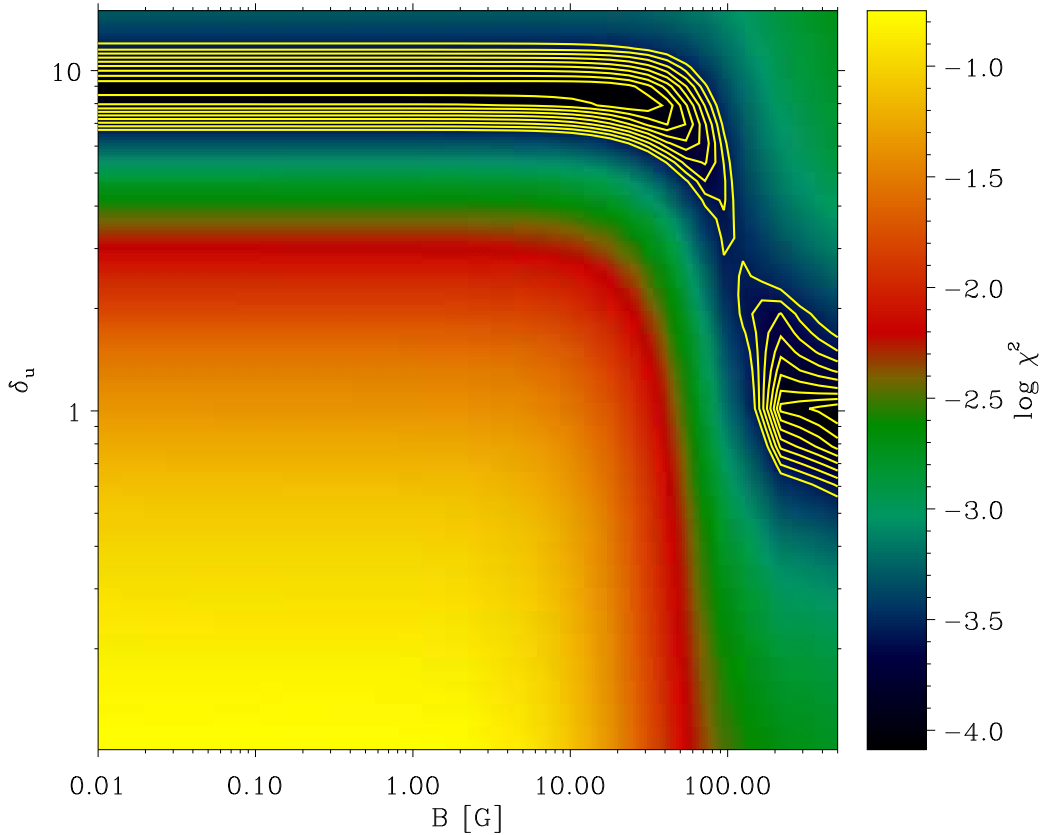}
\caption{The case with collisions and magnetic fields. Values of the $\chi^2$ least-squares function obtained from the
difference between the synthetic and
the observed $Q/I$ signals taking into account the 37 unblended MgH lines that show significant linear polarization amplitudes in \citarwiths{gandorfer_atlas1_00} atlas. We have considered all
possible combinations of the collisional rate ($D^{(2)}$) and of the strength of the microturbulent magnetic field ($B$). As seen in the figure there are two extreme regimes, compatible with the observations, where the $\chi^2$ function reaches the smallest value (i.e., $7.7{\times}10^{-5}$): one with $\langle \delta_u \rangle{\approx}9$ and the other with
$\langle \delta_u \rangle{\approx}0.9$.
We point out that the $\chi^2$-values for the intermediate regime are a factor 3 larger.
Note that the figure gives the impression that the 
($\langle \delta_u \rangle{\approx}9$) 
collisionally dominated solution is compatible with 
magnetic strengths $B{\,}{\lesssim}{\,}30$ G. 
However, the MgH lines with the 
smallest critical Hanle fields indicate that $B{\,}{\lesssim}{\,}10$ G for the collisionally dominated case. 
The weakly collisional solution 
(i.e., that with $\langle \delta_u \rangle{\approx}0.9$)
requires a hidden magnetic field stronger than a few hundred gauss, which we consider unrealistic for the (granular) upflowing regions of the `quiet' Sun.
\label{fig_chisq}}
\end{figure}

Finally, Fig. \ref{fig_chisq} is the most conclusive figure of this Letter, since it summarizes the information we have obtained
from the analysis we have carried out for {\em all} the unblended MgH lines that show sizable $Q/I$ amplitudes in
\citarwiths{gandorfer_atlas1_00} atlas. For each pair of collisional rate ($D^{(2)}$) and of the microturbulent magnetic field strength ($B$), the figure shows the value of the well-known ${\chi}^2$ least-squares function. As expected from the analysis we have done separately for each MgH line (see, e.g., Fig. \ref{fig_qi_field_collis}) we find that (essentially) the following two possibilities lead to an
equally good {\em best fit} to the observed linear polarization amplitudes in {\em all} the unblended 
MgH lines simultaneously: (1) A
collisionally dominated case characterized by $\langle \delta_u \rangle=\langle D^{(2)} \rangle {\,}t_{\rm life} {\approx} 9$ 
and with the
possibility of having a weak microturbulent field whose strength cannot be sensibly larger than 10 G (concerning the case of a
volume-filling and single valued tangled field) and (2) a strongly magnetized case characterized by a microturbulent field of
strength greater than a few hundred gauss, which nevertheless requires the presence of elastic collisions with $\langle \delta_u \rangle = \langle D^{(2)} \rangle {\,}t_{\rm life} {\approx} 0.9$ for fitting the observations\footnote{The calculations we have selected for this section were carried out neglecting inelastic collisions -that is, assuming $\epsilon=0$. However, similar results are obtained for $\epsilon \lesssim 0.5$, and it is very unlikely that the inelastic collisional rates of the MgH lines (which result from electronic transitions whose $A_{ul}{\approx}10^7$s$^{-1}$) are so important to give $\epsilon > 0.5$. In any case, the only significant difference for
$0.5 < \epsilon < 0.8$ lies in the strongly magnetized
solution, which turns out to be characterized by even smaller $\delta_u$-values.}.

\section{Conclusions}

We have shown that in the ``quiet'' regions of the solar photosphere collisions seem to be very efficient in depolarizing
the rotational levels of MgH lines. If the inferred depolarization is 
fully due to elastic collisions, then a single
valued microturbulent field of strength sensibly larger than 10 G filling the whole upflowing volume of the solar photosphere would
be incompatible with the observations. This constraint for the
strength of the hidden field in the (granular) upflowing regions
of the quiet solar photosphere is very gratifying since it is
in agreement with our result that the differential Hanle effect
in the C$_2$ lines of the Swan system implies the presence of very weak fields, with $\langle B \rangle {<} 10$ G
(\citarNP{trujillo_spw3_03}; \citarNP{trujillo_nature04}).

As pointed out by \citar{trujillo_nature04} a volume filling microturbulent field of 10 G is too weak for
explaining the inferred depolarization in the 
(moderately strong) Sr {\sc i} 4607 \AA\ line, whose calculated scattering
polarization amplitude in the absence of magnetic fields turns out to have significant contributions from both the upflowing and
downflowing regions. Actually, when no distinction is made between such regions, the Hanle effect in the strontium line implies
$\langle B \rangle {\approx} 60$ G for the case of a single valued microturbulent field, or $\langle B \rangle {\approx} 130$ G for
the more realistic case of an exponential distribution of field strengths. Trujillo Bueno's et al. (2004) resolution of
this apparent contradiction is that the strength of the hidden field
fluctuates at the spatial scales of the solar granulation pattern, with much stronger fields above the intergranular regions. The above-mentioned constraint, obtained from our analysis
of the scattering polarization in MgH lines, reinforces our previously
reported conclusion that there is a vast amount of hidden magnetic energy and (unsigned) magnetic flux in the inter-network
regions of the `quiet' solar photosphere, carried mainly by rather chaotic fields in the (intergranular) downflowing plasma with
strengths between the equipartition field values and ${\sim}1$ kG (\citarNP{trujillo_nature04}).

Finally, we remark that in this Letter we have also shown that an equally good theoretical fit to the linear
polarization amplitudes observed in MgH lines can be achieved by assuming that the rate of depolarizing collisions is an order of
magnitude smaller than in the previously mentioned collisionally dominated case, but then the required strength of the hidden field
in the (granular) upflowing regions turns out to be unrealistically high. In either case, we find that magnetic fields alone are not
sufficient to explain the observed scattering polarization amplitudes in MgH lines. We are currently investigating whether the inferred collisional depolarization is mainly due to transitions between the Zeeman sublevels of each $J$ level, as we have assumed here for simplicity, or to collisional transitions between different $J$ levels pertaining to the same vibrational and electronic state.

\acknowledgments This research has been funded by the European Commission through the Solar Magnetism Network (contract
HPRN-CT-2002-00313) and by the Spanish Ministerio de Educaci\'on y Ciencia through project AYA2004-05792.


\begin{thebibliography}{17}
\expandafter\ifx\csname natexlab\endcsname\relax\def\natexlab#1{#1}\fi

\bibitem[{{Asensio Ramos} \& {Trujillo Bueno}(2003)}]{asensio_trujillo_spw3_03}
{Asensio Ramos}, A., \& {Trujillo Bueno}, J. 2003, in Solar Polarization 3, ed. J.~{Trujillo Bueno} \& J.~{S\'anchez Almeida}, ASP Conf. Ser. Vol 307, 195

\bibitem[{{Asplund}(2000)}]{asplund00}
{Asplund} M., {Ludwig}, H. G., {Nordlund}, \AA, \& {Stein}, R. F. 2000, \apj, 350, 729

\bibitem[{{Berdyugina} \& {Solanki}(2002)}]{berdyugina02a}
{Berdyugina}, S.~V., \& {Solanki}, S.~K. 2002, \aap, 385, 701

\bibitem[{{Berdyugina} \& {Fluri}(2004)}]{berdyugina_fluri04}
{Berdyugina}, S.~V., \& {Fluri}, D.~M. 2004, \aap, 417, 775

\bibitem[{{Berdyugina} {et~al.}(2002){Berdyugina}, {Stenflo}, \&
  {Gandorfer}}]{berdyugina02b}
{Berdyugina}, S.~V., {Stenflo}, J.~O., \& {Gandorfer}, A. 2002, \aap, 388, 1062

\bibitem[{{Gandorfer}(2000)}]{gandorfer_atlas1_00}
{Gandorfer}, A. 2000, The Second Solar Spectrum, Vol. I: 4625 \AA\ to 6995 \AA\
  (Zurich: vdf)

\bibitem[{{Gandorfer}(2003)}]{gandorfer03}
{Gandorfer}, A. 2003, in Solar Polarization 3, ed. J.~{Trujillo Bueno} \&
  J.~{S\'anchez Almeida}, ASP Conf. Ser. Vol 307, 399

\bibitem[{{Landi Degl'Innocenti}(2003)}]{egidio03}
{Landi Degl'Innocenti}, E. 2003, in Solar Polarization 3, ed. J.~{Trujillo
  Bueno} \& J.~{S\'anchez Almeida}, ASP Conf. Ser. Vol 307, 164

\bibitem[{{Landi Degl'Innocenti} \& {Landolfi}(2004)}]{landi_landolfi04}
{Landi Degl'Innocenti}, E., \& {Landolfi}, M. 2004, Polarization in Spectral
  Lines (Kluwer Academic Publishers)
  
\bibitem[{{MohanRao} \& {Rangarajan}(1999)}]{mohanranga}
{Mohan Rao}, D., \& {Rangarajan}, K.~E. 1999, \apj, 524, L139  
  
\bibitem[{{Stenflo}(2003)}]{stenflo03}
{Stenflo}, J.~O. 2003, in Solar Polarization 3, ed. J.~{Trujillo Bueno} \& J.~{S\'anchez Almeida}, ASP Conf. Ser. Vol 307, 385

\bibitem[{{Stenflo} \& {Keller}(1997)}]{stenflo_keller97}
{Stenflo}, J.~O., \& {Keller}, C.~U. 1997, \aap, 321, 927

\bibitem[{{Trujillo Bueno}(2001)}]{trujillo01}
{Trujillo Bueno}, J. 2001, in Advanced Solar Polarimetry: Theory, Observation
  and Instrumentation, ed. M. {Sigwarth}, ASP Conf.
  Series Vol. 236, 161

\bibitem[{{Trujillo Bueno}(2003{\natexlab{a}})}]{trujillo_spw3_03}
{Trujillo Bueno}, J. 2003{\natexlab{a}}, in Solar Polarization 3, ed.
  J.~{Trujillo Bueno} \& J.~{S\'anchez Almeida}, ASP Conf. Ser. Vol 307, 407

\bibitem[{{Trujillo Bueno}(2003{\natexlab{b}})}]{trujillo03}
{Trujillo Bueno}, J. 2003{\natexlab{b}}, in Stellar Atmosphere Modeling, ed. I.~{Hubeny}, D.~{Mihalas}, \& K.~{Werner}, ASP Conf. Ser. Vol. 288, 551

\bibitem[{{Trujillo Bueno} \& {Manso Sainz}(1999)}]{trujillo_manso99}
{Trujillo Bueno}, J., \& {Manso Sainz}, R. 1999, \apj, 516, 436

\bibitem[{{Trujillo Bueno} {et~al.}(2004){Trujillo Bueno}, {Shchukina}, \& {Asensio Ramos}}]{trujillo_nature04}
{Trujillo Bueno}, J., {Shchukina}, N., \& {Asensio Ramos}, A. 2004, Nature,
  430, 326

\bibitem[{{Weck} {et~al.}(2003){Weck}, {Schweitzer}, {Stancil}, {Hauschildt},
  \& {Kirby}}]{weck03}
{Weck}, P.~F., {Schweitzer}, A., {Stancil}, P.~C., {Hauschildt}, P.~H., \&
  {Kirby}, K. 2003, \apj, 582, 1059

\end{thebibliography}
\end{document}